\documentclass[
    aps,      % American Physical Society journal
    prx,      % PRX Quantum
    reprint,  % approximate final formatting
    noeprint, % exclude eprints from journal references
    nofootinbib, % footnotes at bottom of page, instead of in bib
]{revtex4-2}
\usepackage{appendix}
\usepackage{
    physics,
    mathtools,
    amssymb,
    bm,
    hyperref,
    xcolor,
    graphicx,
}
\graphicspath{{./graphics/drafts1/}}
\usepackage[version=4]{mhchem}
\usepackage{siunitx}
\DeclareSIUnit\angstrom{\text {Å}}
\usepackage{threeparttable}

\usepackage[capitalize]{cleveref}
\usepackage[inline]{enumitem}

\usepackage{booktabs}
\usepackage{csvsimple}

\usepackage{multirow}
\newcommand{\RNum}[1]{\uppercase\expandafter{\romannumeral #1\relax}}

\begin{document}

%\preprint{}
\title{Gatemonium: A Voltage-Tunable Fluxonium}
\author{William M. Strickland}\email[]{wms269@nyu.edu}
\author{Bassel Heiba Elfeky} 
\author{Lukas Baker}
\author{Andrea Maiani}
\author{Jaewoo Lee}
\author{Ido Levy}
\author{Jacob Issokson}
\author{Andrei Vrajitoarea}
\author{Javad Shabani}\email[]{jshabani@nyu.edu}

\affiliation{Center for Quantum Information Physics, Department of Physics, New York University, New York 10003, USA}
\date{\today}

\begin{abstract}
We present a new style of fluxonium qubit, gatemonium, based on an all superconductor-semiconductor hybrid platform. The linear inductance is achieved using six hundred planar Al-InAs Josephson junctions (JJs) in series. By tuning the single junction with a gate voltage, we demonstrate electrostatic control of the effective Josephson energy, tuning the weight of the fictitious phase particle. One and two-tone spectroscopy of the gatemonium transitions further reveal details of the hybrid plasmon-fluxon spectrum. Accounting for the nonsinusoidal current-phase relation of the single junction, we fit the measured spectra to extract charging and inductive energies. We conduct time domain characterization of the plasmon modes and find that energy relaxation times are limited by inductive loss, possibly in the thin aluminum film. We discuss future directions for this platform in gate voltage-tunable, high plasma frequency, enhanced impedance junction arrays, and enhanced coherence times for voltage tunable architectures.
\end{abstract}

%\pacs{}
\maketitle
% Zero point fluctuations of the charge and flux in superconducting circuits give rise to an effective quantum two-level system useful for quantum computation \cite{blais2004}, and more than twenty years ago coherent Rabi oscillations in a superconducting qubit were observed for the first time \cite{Nakamura1999, wallraff_strong_2004}. Since this discovery, there has been an extensive area of research dedicated to improving coherence times of superconducting qubits. One popular strategy is to employ intrinsic error protection through directly engineering the Hamiltonian and resulting wavefunctions. 

\section{Introduction}

Fluxonium qubits have become an incredibly promising alternative to the conventional transmon for quantum computation using superconducting circuits \cite{manucharyan2009,nguyen2022}. The fluxonium computational states have a suppressed charge number transition matrix element when biased to near half a flux quantum, maximizing energy relaxation times \cite{lin2018,earnest2018_heavy, zhang2021}. There also exists a magnetic flux ``sweet spot'', where the fluxonium transition frequency is first-order insensitive to flux noise \cite{hutchings_tunable_2017}. Recently, coherence times exceeding 1 millisecond \cite{somoroff2023} and two-qubit gate fidelities exceeding 99.9\% have been shown with fluxonium qubits \cite{ding2023,bao2022}.

Fundamentally, the fluxonium qubit consists of a Josephson junction (JJ), capacitor, and linear inductor in parallel, with associated energies $E_J$, $E_C$, and $E_L$ respectively. The linear inductance is implemented by a high kinetic inductance element, such as an array of JJs \cite{masluk2012} or a disordered superconductor \cite{hazard2019}. The energy scales $E_J$, $E_C$, and $E_L$ uniquely define the resulting energy spectrum and relevant qubit properties such as frequency and anharmonicity. For example, circuits with $E_J/E_C <1$, in the so-called ``light regime'', have energy levels which become flat with respect to external flux, suppressing flux noise dephasing \cite{Pechenezhskiy2020}. In contrast, circuits with $E_J/E_C>1$, in the so-called ``heavy regime'', have been shown to minimize the charge number transition matrix element near half flux, leading to an enhanced energy relaxation time \cite{earnest2018_heavy, zhang2021, zhang2024}. However, one important point to note is that single mode devices cannot be simultaneously protected from both dephasing and bit flip errors, only one or the other. 

It may be beneficial to \textit{in-situ} tune between the heavy and light regimes, allowing the user to choose whether to maximize bit-flip or phase-flip times at will. Implementations of such a device would necessitate a tunable $E_J$, such as with flux tunable superconducting quantum interference devices (SQUIDs) or voltage-tunable semiconductor junctions \cite{mayer2019}. Devices where the single junction was replaced by a SQUID have been made in the past \cite{lin2018, nguyen2019, Spiecker2023}, however the global magnetic field used to tune $E_J$ also tunes the fluxonium qubit away from the sweet spot. Local flux lines would also cause complications as the SQUID biasing would require calibrations for cross terms in the mutual inductance matrix. Gate-tunable semiconducting junctions could be a useful alternative in this situation, as they would allow for local control over $E_J$ independently of the flux bias. One can also imagine using voltage-tunable junctions in couplers between qubits as detailed in Refs. \citenum{Maxim17} and \citenum{chen2023voltageactivatedparametricentangling} due to the fast, transistor-like voltage-tuning of the semiconductor junction. Recently a tunable inductive coupling was used to implement a two-qubit gate in Ref. \citenum{zhang2024}, and it was shown that flux crosstalk if not carefully canceled can be a leading source of error. The lack of magnetic field crosstalk in voltage-tunable junctions may be beneficial as a coupler and could simplify larger fluxonium architectures \cite{nguyen2022}. 

While voltage-tunable architectures have been shown to have much lower capacitive quality factors ($Q_C = \SI{5e4}{}$ as found in Refs. \citenum{Casparis2018, strickland2024}) than those of state of the art transmons, the possibility of using a fluxonium qubit can actually provide a path forward for enhanced coherence times. For a qubit with frequency $\omega_{01}/2\pi = \SI{65}{\mega Hz}$ and charging energy $E_C/h = \SI{1}{\giga Hz}$, we estimate that due to the lower frequency operation and larger charging energies, gatemonium qubits could have energy relaxation times greater than \SI{10}{\micro s} for inductive quality factors $Q_L = \SI{e4}{}$ (see Appendix A). InAs nanowire weak links have been studied in the past in fluxonium devices such as in Ref. \citenum{pitavidal2020}, but coherent control was not demonstrated. The planar platform also has the distinct advantage that the array can also be made of voltage-tunable junctions, which may be useful for it's larger junction plasma frequency. 

\begin{figure}[h!]
    \centering
    \includegraphics[width=0.45\textwidth]{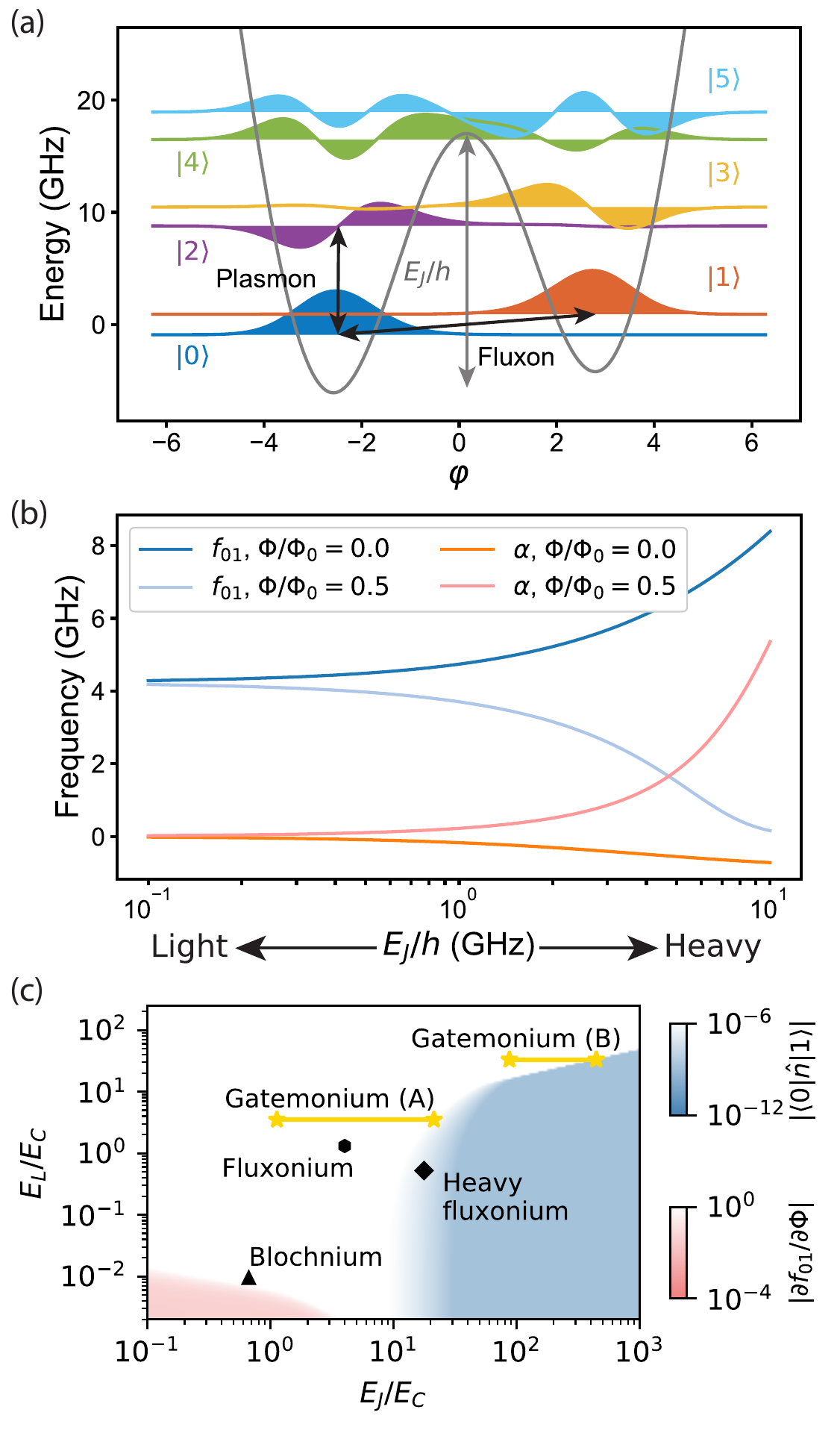}
    \caption{\textbf{$E_J$ tuning in the gatemonium qubit:} (a) Wavefunctions of the lowest levels of the gatemonium qubit in the heavy regime with $E_J/h$ = 17.0 GHz, $E_L/h$ = 2.8 GHz and $E_C/h$ = 0.8 GHz. The external applied flux is $\Phi/\Phi_0 = 0.48$. (b) Qubit frequencies $f_{01}$ and anharmonicities $\alpha$ at zero and half flux. For large values of $E_J$, the qubit frequency at half flux is seen to approach zero and the anharmonicity $\alpha$ becomes greater than 5 GHz. As $E_J$ decreases, moving to the light regime, the spectrum becomes harmonic, as the anharmonicity goes to zero, and the qubit frequency becomes unchanged with flux and approaches $\sqrt{8E_LE_C}/h$. (c) Landscape of fluxonium devices in terms of the relative capacitive, inductive and Josephson energies. We plot the charge number transition matrix element $\langle 0| \hat n | 1\rangle$ at $\Phi/\Phi_0 = 0.45$ in blue color and the derivative of the qubit transition frequency with respect to flux $\partial f_{01}/\partial \Phi$ at $\Phi/\Phi_0 = 0.25$ in red. Relative energies for Fluxonium, Heavy fluxonium, and Blochnium are taken from Refs. \citenum{nguyen2019}, \citenum{earnest2018_heavy}, and \citenum{Pechenezhskiy2020} respectively. The parameter ranges of two gatemonium qubits studied in this paper are also shown in yellow stars from fits to a sinusoidal current-phase relation.}
    \label{fig:theory}
\end{figure}

In this report, we introduce a voltage-tunable fluxonium qubit named ``gatemonium''. The name is a combination of the ``gatemon'', after the gate voltage-tunable transmon nature of the plasmon modes \cite{deLange2015, Larsen15, Casparis2018, yuan2021, strickland2024, wang_coherent_2019, Sagi2024}, and ``fluxonium'', referring to the fluxon modes of a circuit with large shunt inductance. The JJs in the single junction and, more notably, in the array are constructed from superconductor-semiconductor planar JJs. We exploit the \textit{in-situ} $E_J$ control to tune between the heavy and light regimes. As seen in one- and two-tone spectroscopy data, we characterize the gatemonium spectra in terms of the hybrid plasmon and fluxon modes. We fit the energy levels to a nonsinusoidal current-phase relation (CPR) in order to extract the charging and inductive energies. We also demonstrate coherent manipulation of the plasmon mode observed by Rabi oscillations, as well as characterization of the $T_1$ energy relaxation, which is limited by inductive loss. Finally, we discuss the prospects of this qubit platform for higher temperature operation, higher frequency operation, higher impedance arrays, and a voltage-tunable superinductance.

\section{Simulations}

Superconducting quantum circuits operate according to the theory of circuit quantum electrodynamics, where zero point fluctuations of the number of Cooper pairs $\hat n$ and the superconducting phase difference across a Josephson junction $\hat \varphi$ enable macroscopic quantum phenomena \cite{blais2004, Koch2007, blais2021}. The observables $\hat n$ and $\hat \varphi$ form a pair of conjugate variables. In the case of the fluxonium qubit, the circuit can be described by the Hamiltonian
\begin{equation}
    \hat H = 4E_C\hat n^2 + \frac{E_L}{2} \left(\hat \varphi  - 2\pi\frac{\Phi}{\Phi_0} \right) ^2-E_J(\hat \varphi).
\label{eqn:1}
\end{equation}
Each term in the Hamiltonian has an associated energy scale which can be related to physical parameters in the circuit in the following way: the Josephson energy $E_J = \Phi_0 I_C/2\pi$ for a critical current $I_C$, the charging energy $E_C = e^2/2C$ for a capacitance $C$, and the inductive energy $E_L = (\hbar/2e)^2/L$ for an inductance $L$. The magnetic flux quantum is $\Phi_0=h/2e$ with Planck's constant $h$ and the electron charge $e$. Typical for Josephson tunnel junctions, the Josephson potential takes on the form $E_J(\hat \varphi) = - E_J \cos(\hat \varphi)$. However, in general, the energy-phase relation can be written in terms of a sum over the energies of $N$ current carrying channels,
\begin{equation}
    E_J(\hat \varphi)=  \Delta \sum_i^N \sqrt{1- \tau_i \sin^2(\hat \varphi/2)},
    \label{eqn:2}
\end{equation} with superconducting gap $\Delta$ and channel transparency $\tau_i$ for channel $i$ \cite{beenaker1992}. One can recover the conventional cosinusoidal energy-phase relationship in the many-channel, low-transparency limit. An external flux $\Phi$ through the loop tunes the relative depths of the Josephson potential wells. It is often useful to think of the system's classical analog, being a particle with mass $C$ and position $\varphi$ sitting in the bottom of a parabolic potential corrugated by a cosine with amplitude $E_J$.

In Fig. \ref{fig:theory}(a) one can see the resulting wavefunctions in the phase basis of the conventional fluxonium for $E_J/h$ = 17.0 GHz, $E_L/h$ = 2.8 GHz, and $E_C/h$ = 0.8 GHz. The external flux $\Phi$ is set to $0.48 \Phi_0$, close to the half flux degeneracy. The wavefunctions are offset on the y-axis based on their relative energies and labelled accordingly, with the ground state $|0\rangle$ in blue, the first excited state $|1\rangle$ in orange, and so on. The potential energy is shown in gray with the barrier height between the lowest two wells being $E_J/h$. One may begin to notice two distinct transitions from the ground state: transitions within each well corresponding to plasmon modes, and transitions between different wells corresponding to fluxon modes. Relevant to qubit operation, we plot the exact values of the lowest energy transition frequency $f_{01}$ (blue) and anharmonicity $\alpha$ (red) as a function of $E_J$ in Fig. \ref{fig:theory}(b) at zero (dark color) and half (light color) flux. It can be seen that at large $E_J$, the qubit frequency at half flux approaches zero, while the anharmonicity exceeds 5 GHz. As $E_J$ decreases to less than $E_C$, the qubit spectrum becomes harmonic, leading to a vanishing anharmonicity. In addition, the qubit becomes less flux tunable, as inferred from the difference in frequency between zero and half flux. The frequency approaches a value set by $\sqrt{8E_LE_C}/h$. 

To reiterate the importance of the different fluxonium parameter regimes, we plot the $E_J$, $E_L$, and $E_C$ landscape in Fig \ref{fig:theory}(c), adapted from Ref. \citenum{gyenis2021}. Experimentally realized device parameters are shown on the plot \cite{nguyen2019, earnest2018_heavy, Pechenezhskiy2020}. We plot both the charge number transition matrix element $\langle 0|\hat n |1\rangle$, calculated at $\Phi/\Phi_0 = 0.45$, and the derivative of the $\partial f_{01}/\partial \Phi$, calculated at $\Phi/\Phi_0 = 0.25$. We also plot the achievable range of the two gatemonium devices studied in this paper. The values are calculated from the measured plasmon frequency at zero flux (see Appendix B). While it may be advantageous to be in the heavy regime for bit-flip protection and the light regime for phase flip protection, we show that with our gatemonium device we are able to tune the ratio of $E_J/E_C$ by more than an order of magnitude, connecting these two regimes. 

% The Josephson potential localizes wavefunctions to individual wells, and the barrier height along with $E_L$ determines the suppression of the transition matrix element, influencing energy relaxation times $T_1$. One can engineer a ``heavy'' fluxonium, where $E_J > E_C$, in order to maximize energy relaxation times, as was done in Refs. \citenum{lin2018, earnest2018_heavy, zhang2021}. Also interesting is to consider the susceptibility to flux noise through the derivative of the $|0\rangle$ to $|1\rangle$ transition frequency with respect to flux $df_{01}/d\Phi$. In the ``light'' regime, where $ E_J < E_C$, the qubit is insensitive to flux, and hence more protected from flux noise dephasing as shown in Ref. \citenum{Pechenezhskiy2020}. This principle also illustrates the advantages of working at the half flux sweet spot, since the qubit is first-order insensitive to flux noise. A more detailed description of the error protection properties of fluxonium can be found in Ref \citenum{gyenis2021}. 

\begin{figure}[t!]
    \centering
    \includegraphics[width=0.45\textwidth]{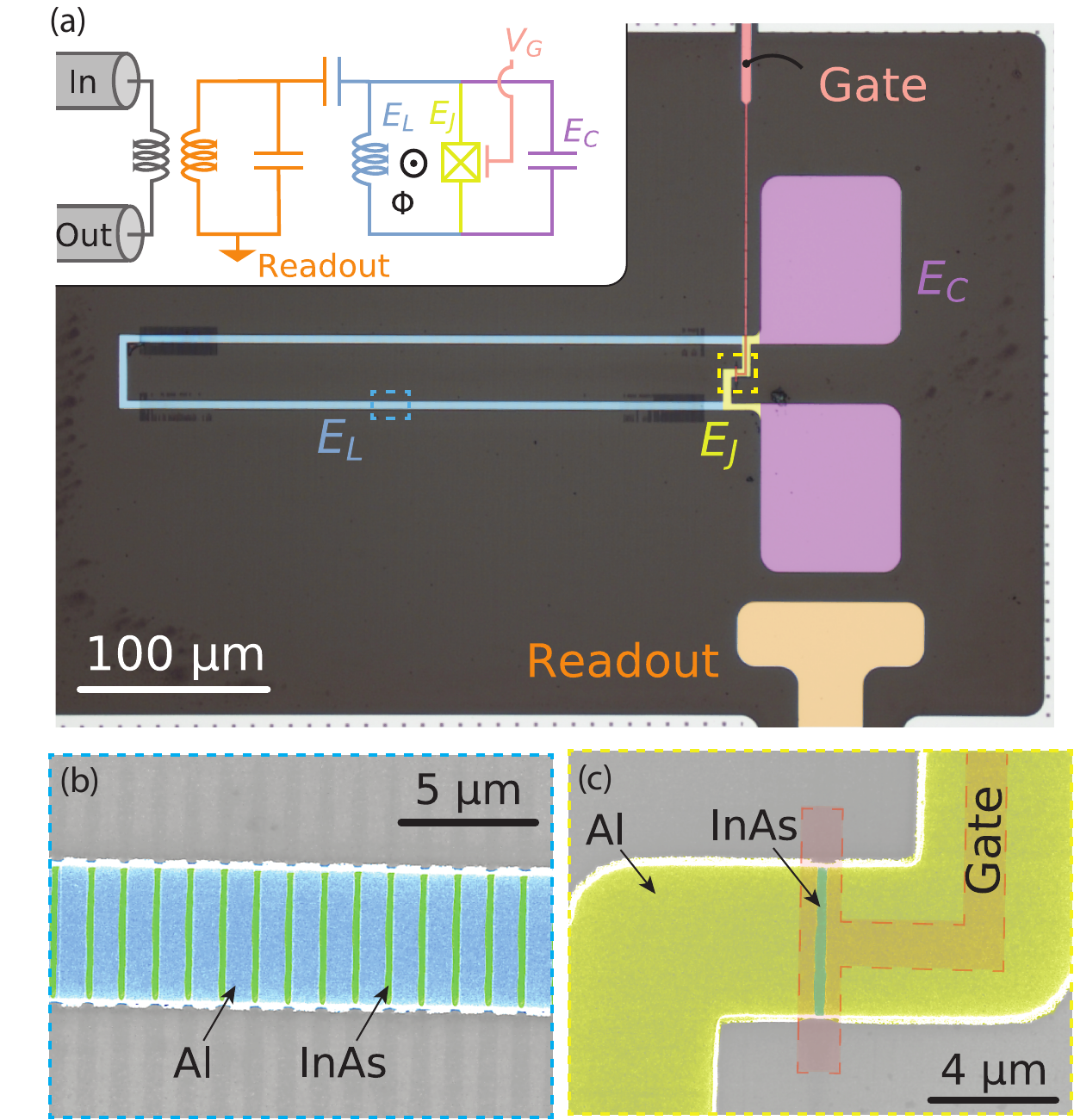}
    \caption{ \textbf{Device A optical image and circuit diagram:} False-colored optical image of the gatemonium qubit is shown in (a) consisting of a shunt capacitor with charging energy $E_C$ (purple), a planar JJ with Josephson energy $E_J$ (yellow), and a linear inductance, implemented in the form of a series of planar JJs with inductive energy $E_L$ (blue). The inset shows an equivalent circuit diagram. Input and output transmission lines are inductively coupled to a readout resonator, which is capacitively coupled to the gatemonium. A zoomed in scanning electron micrograph (SEM) is shown in (b) of the JJ array, with aluminum islands (blue) connected via Josephson coupling through InAs weak-links (green). An SEM image of the single junction before gate deposition is shown in (c) with aluminum leads (yellow) separated by an InAs weak link (green) and the top gate placement (red).}
    \label{fig:fab}
\end{figure}

\section{Device design and fabrication}

We discuss two devices, A, and B, with their design and extracted qubit parameters shown in Table \ref{table:1}. The growth and fabrication details of these devices can be found in Appendix C. We show a false-color optical image of Device A in Fig. \ref{fig:fab}(a) with the capacitor in purple, the single JJ in yellow, the array inductor in blue, the readout resonator in orange, and the gate in red. The equivalent circuit diagram is shown in the inset with colors corresponding to the optical image. The array is implemented by several hundred planar JJs in series, where a false-color scanning electron micrograph (SEM) is shown in Fig. \ref{fig:fab}(b). Al islands (blue) 1$\times$5 \SI{}{\micro m}$^2$ in size are connected through InAs weak links (green). A false color SEM image of the single junction is shown in Fig. \ref{fig:fab}(c) before gate deposition, with Al leads in yellow and the InAs region in green. The top gate electrode is used to tune $E_J$, and shown schematically in red. The bottom qubit capacitor pad is coupled to a readout resonator with a coupling strength of $g/2\pi = \SI{150}{\mega Hz}$ as measured from the minimum detuning of the vacuum Rabi splitting \cite{wallraff_strong_2004}. The readout resonators are $\lambda/4$ coplanar waveguides with frequencies and coupling rates to the feedline shown in Table \ref{table:1}. The internal quality factor of Device A is \SI{5.5e3}{} measured at low power with the qubit far detuned and is consistent with previous devices that underwent similar device fabrication \cite{strickland_superconducting_2023}. Complex transmission across the feedline $S_{21}$ is measured as a function of probe frequency $f_\mathrm{probe}$. We also note that an external charge line coupled to the qubit capacitor.

The device chip is mounted in a BeCu sample holder and shielded by aluminum and Cryoperm cans. Transmission lines on the chip are bonded to a printed circuit board using Al wirebonds. The package is mounted at the mixing chamber plate of a Oxford Instruments Triton, a cryogen-free dilution refrigerator, with a base temperature of \SI{12}{\milli K}. Using a coil on the back of the sample holder we apply a global external magnetic flux through the loop $\Phi$. A schematic of the wiring is shown in Appendix D.

\begin{table*}[t]
\centering
\begin{tabular}{|| c | c | c | c | c | c| c ||} 
 \hline
 Device & $f_r$ (GHz) & $\kappa/2\pi$ (MHz) & $N_{JJ}$ & $W_C (\SI{}{\micro m}$) & $E_L/h$ (GHz) & $E_C/h$ (MHz) \\ [0.5ex] 
 \hline\hline
  A & 7.41 & 1.95 & 600 & 75.0 & 2.80 & 800 \\
  B & 7.09 & 0.77 & 400 & 500 & 4.97 & 150 \\ %JS501
  % C & 400 & 75 & 4.65 & 800 \\ %JS501
  \hline
\end{tabular}
\caption{\textbf{Device parameters} of Devices A, and B with the readout resonator frequency $f_r$, the coupling to the feedline $\kappa/2\pi$, the number of JJs in the array $N_{JJ}$, the width of the capacitor $W_C$, as well as the inductive energy $E_L$, and the charging energy $E_C$ determined by fits to two-tone spectroscopy data.}
\label{table:1}
\end{table*}

\section{Resonator and qubit spectroscopy}

\begin{figure}[t!]
    \centering
    \includegraphics[width=0.5\textwidth]{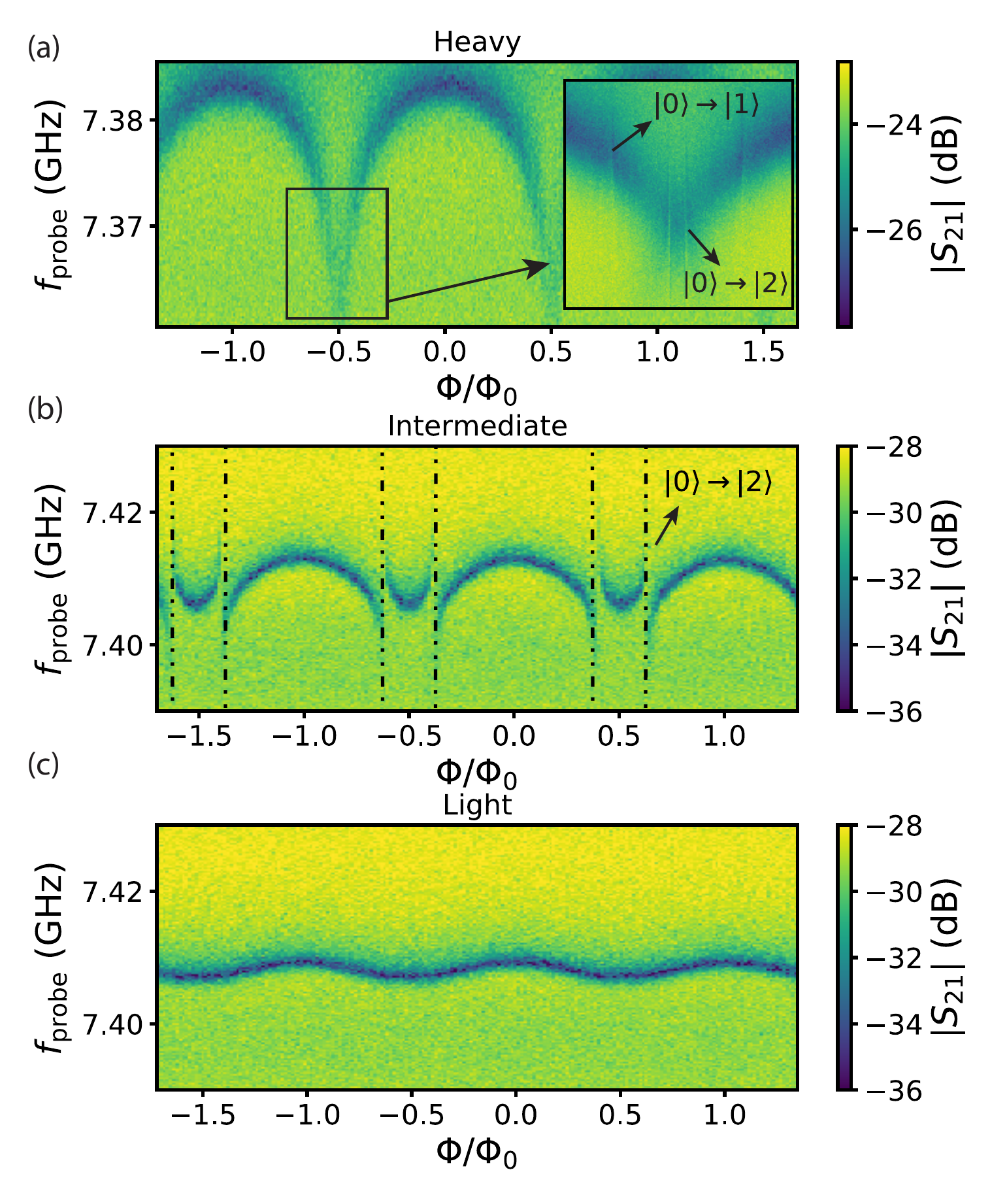}
    \caption{\textbf{Resonator spectroscopy of Device A:} Transmission across the feedline $|S_{21}|$ is shown as a function of applied external magnetic flux $\Phi/\Phi_0$ at different gate voltages. (a) The gatemonium is in the heavy regime and the dressed resonator frequency tunes periodically with flux due to coupling with the plasmon mode. Zooming in to the half flux (inset), one can see multiple crossings with the qubit $f_{01}$ and $f_{02}$. (b) In the intermediate regime the fluxon and plasmon modes become hybridized and we see $f_{02}$ cross with the readout resonator near half flux. (c) In the light regime, the detuning of the gatemonium mode with the resonator increases, leading to a less pronounced flux dependence.}
    \label{fig:onetone}
\end{figure}

\begin{figure*}[t!]
    \centering
    \includegraphics[width=\textwidth]{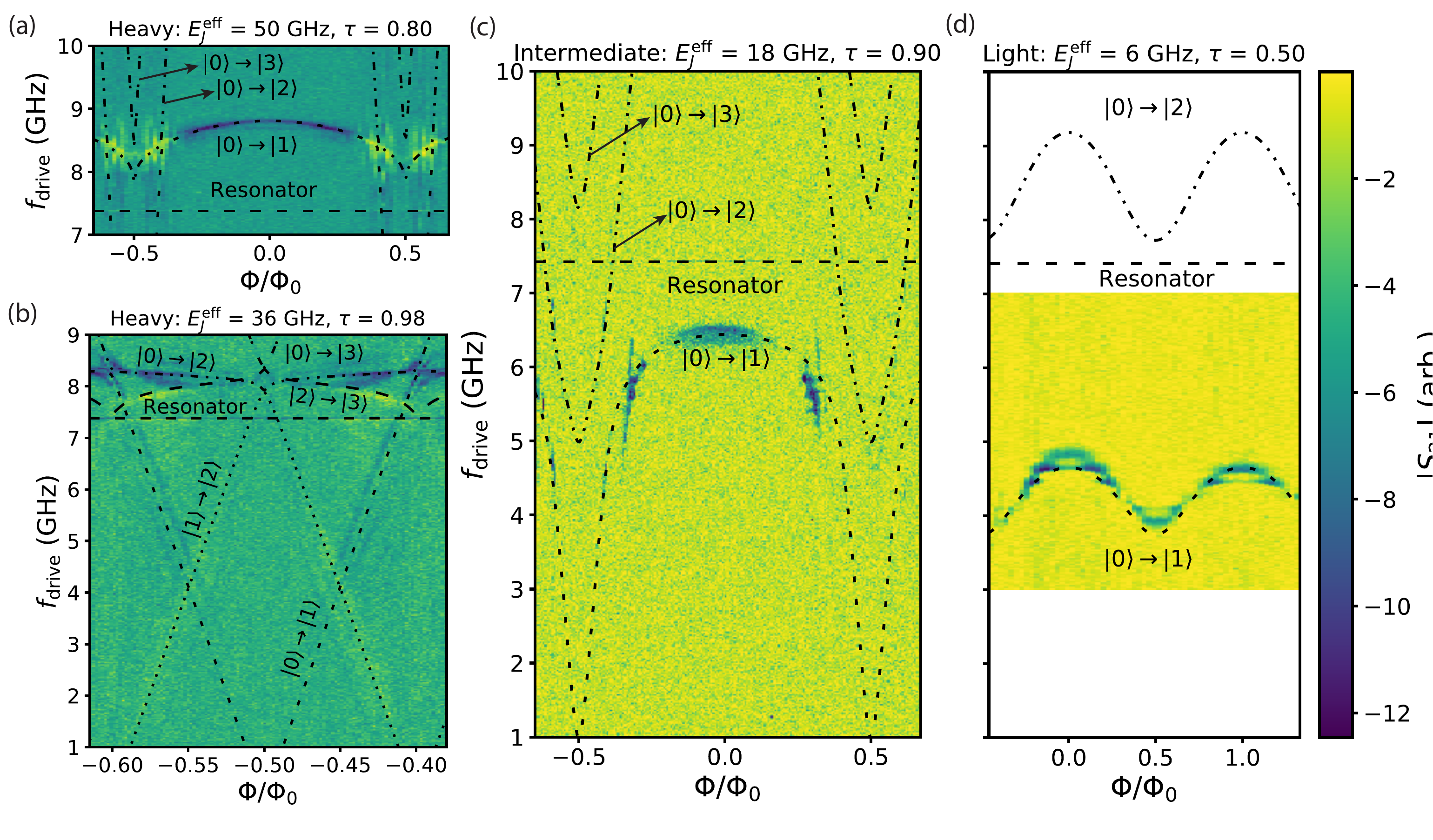}
    \caption{\textbf{Two-tone spectroscopy of Device A:} (a) In the heavy regime $E_{J}^\mathrm{eff}/h = \SI{50}{\giga Hz}$ the plasmon is weakly flux tunable and exhibits multiple avoided crossings near half flux. (b) Zooming in to near half flux, one can see the linearly dispersing fluxon mode, as well as various other transitions, which are labelled. (c) For $E_{J}^\mathrm{eff}/h$ = 18 GHz the fluxon and plasmon modes are hybridized with $f_{01}$ reaching near \SI{1.0}{\giga Hz} at half flux and \SI{6.2}{\giga Hz} at zero flux. (f) As $E_{J1}$ decreases, the gatemonium spectrum becomes harmonic and weakly flux tunable, as seen by the sinusoidal behavior of $f_{01}$ with flux, centered around a value of $\sqrt{8E_LE_C}$. Asymmetry with respect to flux can be attributed to drift in  the qubit frequency over the course of the measurement (see Appendix F).}
    \label{fig:twotone}
\end{figure*}
We present qubit spectroscopy at different effective $E_J$ values. The Josephson energy is tuned via an applied top gate voltage, with the exact tuning curve shown in Appendix B. With the qubit in the heavy regime, we measure $|S_{21}|$ across the feedline while sweeping $f_\mathrm{probe}$ near the readout resonator frequency, shown in Fig. \ref{fig:onetone}(a). The qubit plasmon transition frequency at zero flux is detuned by more than 1 GHz above the resonator frequency as determined by two-tone spectroscopy. The resonator is weakly tuned with flux due to dressing of the resonator frequency by the flux tunable plasmon modes. Upon zooming in to $\Phi/\Phi_0=0.5$ (inset), one finds multiple weak avoided crossings between the resonator and qubit modes. As we will learn later by fitting two-tone spectroscopy data seen in Figs. \ref{fig:twotone}(a-c), these transitions correspond to the $|0\rangle$ to $|1\rangle$ and $|0\rangle$ to $|2\rangle$ transitions of the qubit. As the gate voltage decreases, the gatemonium qubit approaches the intermediate regime, between the heavy and light regimes. In Fig. \ref{fig:onetone}(b) the resonator exhibits a strong avoided level crossing with the qubit mode while still tuning with flux. Here the qubit's plasmon transition at zero flux is about 1 GHz detuned below the resonator. We then tune further into the light regime in Fig. \ref{fig:onetone}(c), where the resonator is now very weakly flux tunable. This is in agreement with the observation made in Fig. \ref{fig:theory}(b), where as the gatemonium $E_J$ approaches zero the flux tunability of the qubit frequency decreases. The qubit transition at this point oscillates around 4 GHz with flux.

% An $E_L = \SI{2.8}{\giga Hz}$ corresponds to a $\SI{1.15}{\micro H}$ linear inductance.
% Operating at a frequency of 4.2 GHz this would give a characteristic impedance of $\SI{4.84}{\kilo \Omega}$, very close to the superconducting resistance quantum $\SI{6.45}{\kilo \Omega}$. 

Understanding the effect of the qubit on the dressed resonator frequency, we now probe the qubit transitions directly by conducting two-tone spectroscopy. We apply a drive tone with varying frequency $f_\mathrm{drive}$ on either the external charge line or the gate line, and again measure $|S_{21}|$ across the feedline near the frequency of the readout resonator. Excitations of the qubit cause a state-dependent shift of the resonator frequency when the qubit is excited, allowing one to measure the qubit response \cite{blais2004}. With the gatemonium once again in the heavy regime, we conduct two-tone spectroscopy, where the results are shown in Fig. \ref{fig:twotone}(a). One can see a transition which tunes with flux at 8.9 GHz at zero flux to 8.0 GHz at half flux. In order to properly understand the gatemonium spectrum, we solve for the expected transition frequencies for the Hamiltonian shown in Eq. \ref{eqn:1}. We assume that the Josephson potential can be accurately described by a single conduction channel with average transparency $\tau$, 
\begin{equation}
    E_J (\hat \varphi) \approx E_J^\mathrm{eff} \sqrt{1 - \tau\sin^2\left(\frac{\hat \varphi}{2}\right)}.
\end{equation}
where the energy scale $E_J^\mathrm{eff}$ takes into account the superconducting gap, and the number of channels. With this simplification, we plot the theoretical transition frequencies on top of the measured spectra and find good agreement for $E_J^\mathrm{eff}$ = 50 GHz and $\tau = 0.80$. We find that the transition observed in Fig. \ref{fig:twotone}(a) corresponds to a plasmon mode, and some avoided level crossings with fluxon modes near $\Phi/\Phi_0 = 0.40$ are visible. In Appendix E, we also show fits to the data using a sinusoidal CPR with a comparison between the two models. 
 % we plot the theoretical transition frequencies on top of the measured spectra and find good agreement, seen in orange and pink dashed lines, for $E_{J1}/h = \SI{12.0}{\giga Hz}$, $E_L/h = \SI{2.80}{\giga Hz}$  and $E_C/h$ = 0.80 GHz in Fig. 1(a). 
 
In order to resolve the fluxon modes more clearly, we zoom in to near half flux and measure to much lower drive frequencies after decreasing the gate voltage, shown in Fig. \ref{fig:twotone}(b). We see that as the flux increases towards $\Phi/\Phi_0 = 0.50$, the fluxon mode disperses linearly towards low frequency. One can also notice other transitions, such as the $|1\rangle$ to $|2\rangle$, $|0\rangle$ to $|2\rangle$, $|0\rangle$ to $|3\rangle$, and $|2\rangle$ to $|3\rangle$. In order to couple to the fluxon transitions we employ a higher power drive, which may cause resonator-induced leakage \cite{sank2016}. The qubit frequency at half flux in this regime reaches as low as $f_{01}$ = 75 MHz inferred from the fit. Again we see the data fit well to theory for $E_J^\mathrm{eff}$ = 36 GHz and $\tau = 0.98$. We find that the transparency increases as $E_J^\mathrm{eff}$ goes down from Fig. \ref{fig:twotone}(a) to (b), and may be related to the nonmonotonic tuning with gate voltage, detailed in Appendix B. Dependence of the transparency as a function of gate voltage in an InAs planar JJ was shown recently shown in Ref. \citenum{elfeky2024microwaveandreevboundstate} and exhibited a similar behavior.
 
We next tune to the intermediate regime, shown in Fig. \ref{fig:twotone}(c), where the plasmon mode at zero flux now has a frequency of 6.2 GHz, and tunes to around 1 GHz at half flux. The $|0\rangle$ to $|2\rangle$ transition exhibits an avoided crossing with the readout resonator at 7.4 GHz, yielding a response similar to that seen in Fig. \ref{fig:onetone}(b). One can see that the plasmon and fluxon modes in the intermediate regime are much more strongly coupled as compared to those in the heavy regime. Fitting the data we see these transitions correspond to a spectrum with $E_J^\mathrm{eff}$ = 17 GHz and $\tau$ = 0.90. The gate voltage is then decreased such that we enter the light regime, where the qubit response to flux is nearly sinusoidal, centered around a frequency of $4.2$ GHz, shown in Fig. \ref{fig:twotone}(d). In this case the spectrum is purely harmonic with a lowest transition frequency equal to $\sqrt{8E_LE_C}$. Here we find that $E_J^\mathrm{eff}$ = 6 GHz and $\tau$ = 0.50. We note that over the course of the measurement, the value of the Josephson energy can drift over time due to $1/f$-type noise on the gate voltage, which lead to an asymmetry in qubit response as a function of magnetic flux. We detail this further in Appendix F. 

Since gate-tuning the single junction should only have an effect on the effective Josephson energy, the values of $E_L$ and $E_C$ should be consistent whether the gatemonium qubit is in the heavy or light regime. Indeed this is the case, where all the theory curves fit the data with $E_L/h$ = 2.80 GHz and $E_C/h$ = 0.80 GHz. The inductive energy corresponds to an inductance of 58 nH in the array. This yields a critical current per junction of about \SI{3.3}{\micro A}, consistent with similar planar InAs junctions \cite{mayer2019}. In the future we would like to enhance this inductance by narrowing the size of the junctions, effectively reducing the number of current carrying modes, and by adding more junctions to the array.

\begin{figure}
    \centering
    \includegraphics[width=0.45\textwidth]{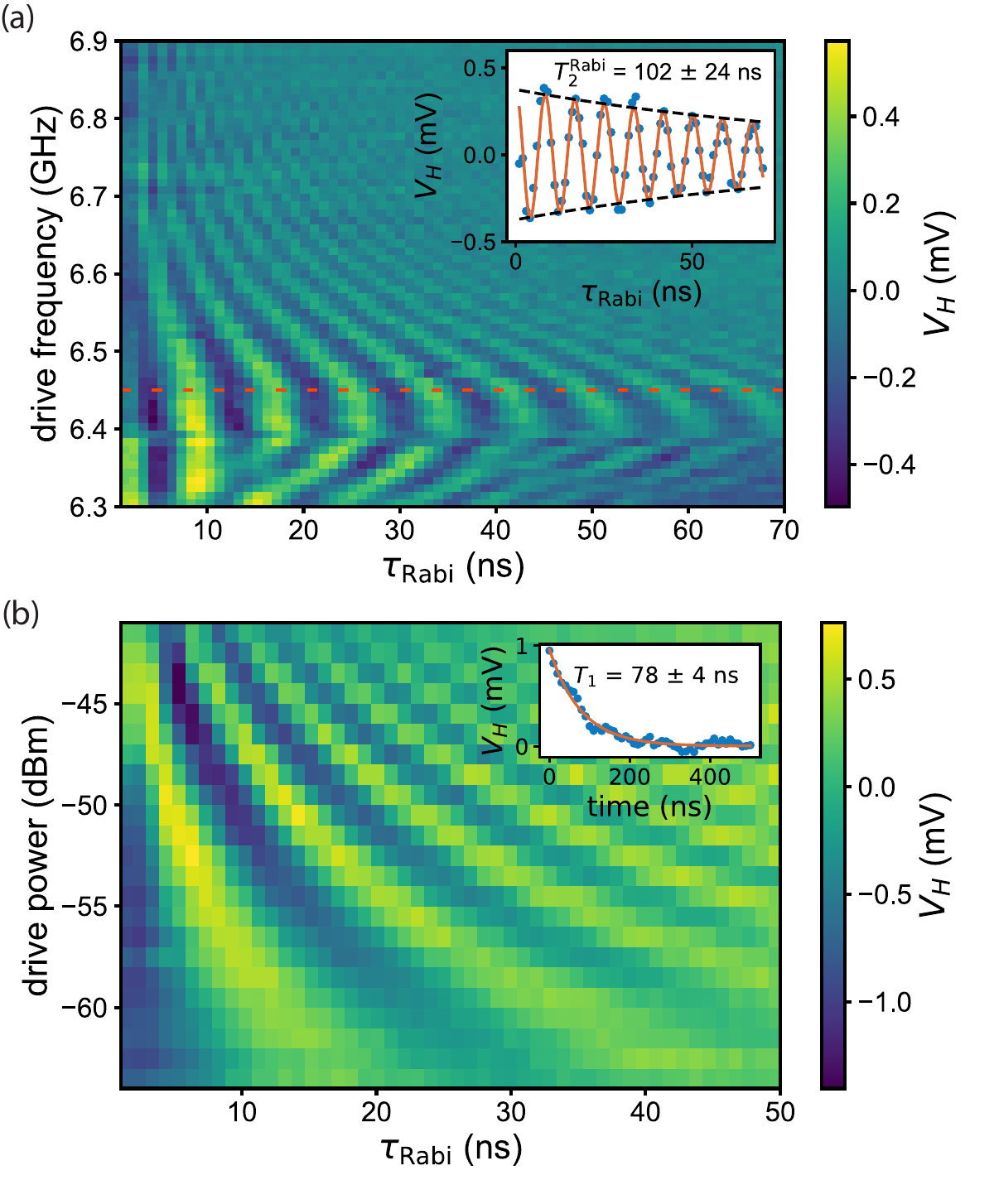}
    \caption{\textbf{Time domain measurements of Device B in the plasmon regime}: Rabi oscillations as a function of drive frequency and drive power are shown in (a) and (b) respectively. A linecut as a function of pulse width $\tau_\mathrm{Rabi}$ for a fixed drive frequency of 6.45 GHz is shown in the inset (blue points), with a fit to the decaying sinusoid shown (yellow solid line with black dotted line the decay envelope) yielding $T_2^\mathrm{Rabi} = \SI{102}{\nano s}$. }
    \label{fig:timedomain}
\end{figure}

\section{Plasmon time domain characterization}
\label{section:timedomain}
After properly characterizing the fluxonium spectra, we employ homodyne detection of the qubit state in a variety of pulsed measurements on Device B. This device has a much large capacitance than Device A, with a charging energy of 150 MHz. Over all gate voltages the qubit is in the heavy regime. By measuring the spectrum near zero flux, we set the flux to $\Phi/\Phi_0 = 0.0$ such that the lowest energy transition corresponds to a plasmon transition for $f_{01}$ = 6.50 GHz. Two tone spectroscopy at various gate voltages for Device B are shown in Appendix G.

We apply a square pulse near the qubit frequency with a width of $\tau_\mathrm{Rabi}$ while applying a weak continuous readout. Results are shown in Fig. \ref{fig:timedomain}(a) for a varying drive frequency. The data shown is integrated over the entire 200 ns readout duration. We find that as a function of $\tau_\mathrm{Rabi}$, the homodyne detection voltage $V_H$ oscillates, corresponding to Rabi oscillations. It can be seen that the Rabi frequency changes as a function of drive detuning from the qubit frequency, being 6.42 GHz, resulting in an expected ``chevron'' pattern. The inset shows a linecut of the data at a fixed $f_\mathrm{drive}$ = \SI{6.45}{\giga Hz} shown in blue points. The data is fit to a decaying sinusoid, shown in the orange solid line, revealing the coherence time associated with the Rabi manipulation, being 102 $\pm$ 24 \SI{}{\nano s}. 
The decay envelope is shown with a black dotted line. We note that the feature near $f_\mathrm{drive}$ = 6.40 GHz is due possibly to fluctuations of the Josephson energy, which we detail in Appendix F. 
We also find the Rabi frequency decrease as we decrease the drive power as shown in Fig. \ref{fig:timedomain}(b). We apply a pulse with width equal to half the Rabi period in order to drive a rotation about the Bloch sphere of an angle $\pi$, exciting the gatemonium from the $|0\rangle$ to the $|1\rangle$ state. By varying the readout time after the $\pi$-pulse, we are able to fit the decay of the qubit, yielding an energy relaxation time $T_1 = 78 \pm 4$ ns, shown in the inset. 
% This is consistent with gatemon $T_1$ times measured on using a similar fabrication procedure \cite{strickland2024}, and is most likely limited by inductive loss in the thin epitaxial Al.

From this measurement of the energy relaxation time in Device B, we would like to estimate the inductive quality factor of the array $Q_L$. Details of the calculation are included in Appendix A. We assume a capacitive quality factor $Q_C = \SI{1.2e4}{}$, measured in CPW resonators limited by the blanket gate dielectric. For the energy scales of Device B shown in Table \ref{table:1}, we are able to solve for the expected $T_1$ following the expressions in Eqs. \ref{eqn:inductive} and \ref{eqn:capacitive} assuming only inductive and capacitive loss. We find that $T_1 = \SI{78}{\nano s}$ and $Q_C = \SI{1.2e4}{}$ corresponds to an inductive quality factor of $Q_L \sim \SI{630}{}$, consistent with measurements of the inductive loss of CPWs with thin aluminum films in the large kinetic inductance limit (see Ref. \citenum{strickland2024}). We also find the capacitive and inductive dissipation rates are $\Gamma_\mathrm{cap} \sim \SI{3}{\mega Hz}$ and $\Gamma_\mathrm{ind} \sim \SI{20}{\mega Hz}$ respectively, thus the energy relaxation time is limited primarily by inductive loss. We believe that increasing the thickness of the aluminum could enhance inductive quality factors. To disambiguate losses in the aluminum film and the array of junctions, lumped element resonators similar to those found in Ref. \citenum{masluk2012} can be measured. A comprehensive study of the loss of the junction would also be quite useful in understanding the limits to inductive loss in InAs junction arrays and should be the topic of future study.

With a very small charging energy of 150 MHz it is difficult to couple to fluxon modes with a capacitively coupled charge line. It could be possible to utilize a Raman driving procedure as used in Ref. \citenum{earnest2018_heavy} to couple to the fluxon modes in the heavy regime. In a future experiment we would like to measure the coherence properties of the fluxon modes in the gatemonium qubit. It would also  be interesting to investigate the effect of quasiparticle \cite{elfeky_quasiparticle2023} and phase slip processes \cite{randeria2024} on gatemonium dephasing.

% We discuss the coherence properties of the gatemonium device. We know the phase slip rate in the array junctions to be proportional to $\exp[-(8E_{JA}/E_{CA})^{1/2}]$ for array junction Josephson and charging energies of $E_{JA}$ and $E_{CA}$ respectively. Because of the nature of the coplanar capacitance in our junctions, we believe this gives rise to large $E_C$. Furthermore, $E_{JA}$ = 1.680 THz. This could lead to an enhanced phase slip rate, decreasing the relative dephasing. We also find that the 

% for discussion on how array modes couple to plasmon/fluxon modes look at Roman Kuzmin Nature 2019

\section{Conclusion}

We have presented a thorough introduction to the gatemonium qubit, where the inductor was achieved using 600 superconductor-semiconductor JJs. We provide a detailed analysis of the gatemonium spectra at different effective $E_J$ regimes as a function of flux, consistent with a nonsinusoidal CPR. 
The gatemonium $E_J$ is tunable, accessing both the light and heavy fluxonium regimes. 
We also present measurements of the coherence times from Rabi oscillations and a $T_1$ measurement of the plasmon mode in the heavy regime. We find the energy relaxation times are limited by inductive loss for an inductive quality factor $Q_L \sim 630$, consistent with thin Al films with large kinetic inductance. 

We believe this is an exciting first step to making fluxonium-style qubits on this material platform. Recently there has been increased interest in using different types of materials for high kinetic inductance junction arrays. The use of superconductor-semiconductor planar junctions can lead to a higher Josephson plasma frequencies, possibly for higher operating temperatures and for achieving larger inductances. While the junction leads in conventional superconductor-insulator-superconductor junctions effectively form a parallel plate capacitor, the junction leads in a superconductor-semiconductor junction form a coplanar capacitor, yielding a reduced Josephson capacitance $C_J$ and increased Josephson plasma frequency $\omega_p = 1/\sqrt{L_JC_J}$ for $L_J$ the Josephson inductance. Since the maximum operating frequency of the array is enhanced, one can imagine operating fluxonium qubits at higher temperatures based on these materials. 
In addition, it was found in Ref. \citenum{vanevic2012} that the phase slip rate in an array of highly transparent junctions may be exponentially suppressed due to a renormalization of the charging energy. While the phase slip rate in conventional aluminum oxide junction arrays increases with junction impedance, this exponential suppression can actually help enable higher impedance arrays. We also note that the semiconductor weak-links in the array can give rise to a voltage-tunable superinductance \cite{Bell2012}. For these reasons, a JJ array based on superconductor-semiconductor hybrid materials may be useful for superconducting qubits, amplifiers \cite{phan2022, Hao2024, splithoff2024}, couplers \cite{sardashti2020, materise2023, Maxim17, Casparis2019} and metamaterials \cite{splitthoff2024gatetunable} in the future.

\section{Data availability}
The data that support the findings of this article are publicly available (\href{https://doi.org/10.5281/zenodo.13800028}{https://doi.org/10.5281/zenodo.13800028}) \cite{Strickland2024-ly}. All code pertaining to the work is available from the corresponding author upon request.

\section{Acknowledgments}
We thank Andrew P. Higginbotham, Shyam Shankar, Archana Kamal, Maxim G. Vavilov, Vladimir E. Manucharyan, Srivatsan Chakram, Peter Schüffelgen, and Charles Tahan for fruitful conversations. We also acknowledge Andras Gyenis for providing code to produce the plot in Fig. \ref{fig:theory}(c). We acknowledge support from the Army Research Office under agreement W911NF2110303 and from the DARPA Synthetic Quantum Nanostructures (SynQuaNon) program under agreement HR00112420343. W.M.S. acknowledges funding from the ARO/LPS QuaCR Graduate Fellowship. The authors acknowledge MIT Lincoln Laboratory (LL) and IARPA for providing the traveling wave parametric amplifier used in this work. We also acknowledge the MIT LL SQUIL Foundry for a transmon qubit device to help with testing fridge wiring. 

\bibliography{references_shabani_growth}
\newpage
\clearpage

\begin{figure}
    \centering
    \includegraphics[width=\linewidth]{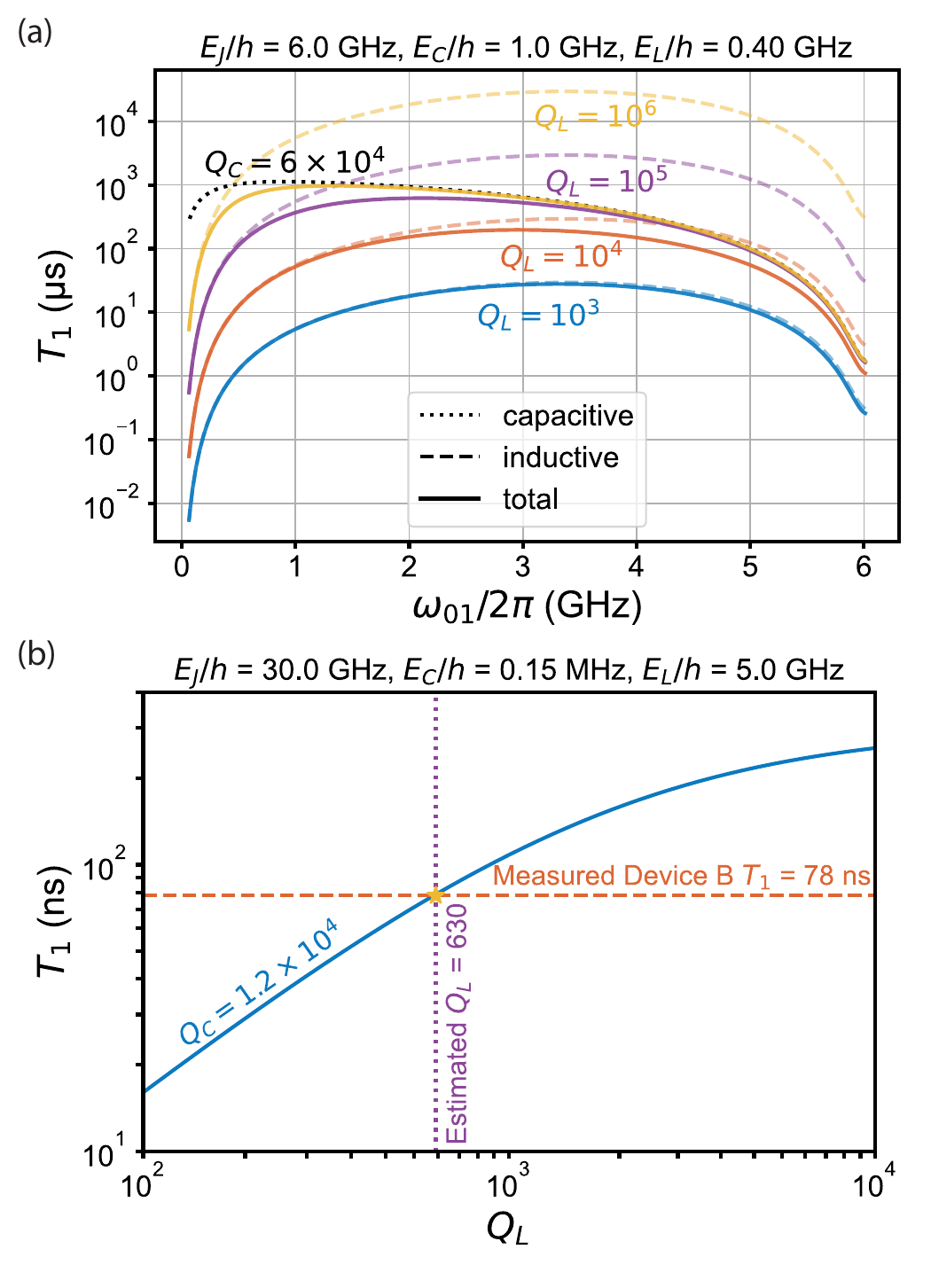}
    \caption{\textbf{Estimated $T_1$ and $Q_L$ of a gatemonium qubit.} (a) As a function of qubit frequency $\omega_{01}$ with $E_J/h = \SI{6}{\giga Hz}$, $E_C/h = \SI{1}{\giga Hz}$, $E_L/h = \SI{0.4}{\giga Hz}$, and temperature $T = \SI{12}{\milli K}$ we plot the $T_1$ limits due to capacitive loss (dotted black line) for a capacitive quality factor of $Q_C = \SI{6e4}{}$, and inductive loss for different inductive quality factors, being $Q_L$ = \SI{1e3}{}, \SI{1e4}{}, \SI{1e5}{}, and \SI{1e6}{} (blue, orange, purple, and yellow dashed lines respectively). The total $T_1$ is the inverse of the sum of the dissipation rates $(\Gamma_\mathrm{ind} + \Gamma_\mathrm{cap})^{-1}$ and is shown in solid lines. (b) Estimate of $Q_L \sim 630$ for Device B with $T_1 = \SI{78}{\nano s}$ at zero flux, corresponding to a frequency $\omega_{01}/2\pi = \SI{6.35}{\giga Hz}$. We estimate the $Q_C = \SI{1.2e4}{}$, and for measured energy scales $E_J/h = \SI{30.0}{\giga Hz}$, $E_C/h = \SI{0.15}{\giga Hz}$, and $E_L/h = \SI{5.0}{\giga Hz}$.}
    \label{fig:t1_estimate}
\end{figure}

\begin{figure}[t!]
    \centering
    \includegraphics[width=0.5\textwidth]{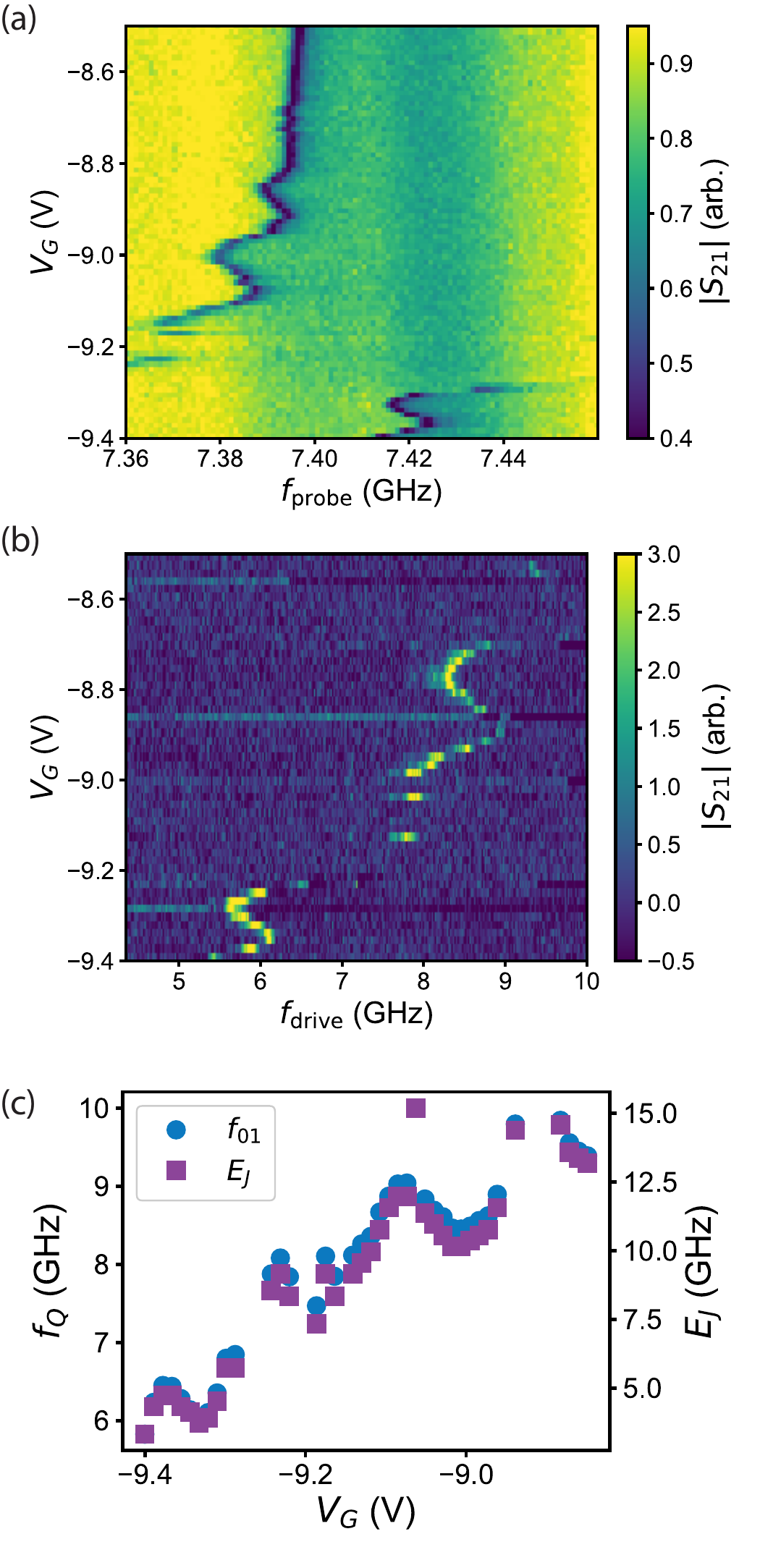}
    \caption{ \textbf{Gate voltage tuning of $E_J$ in Device A}: (a) The top gate voltage $V_G$ tunes the critical current in the JJ, forcing the qubit mode to undergo an avoided level crossing with the readout resonator. (b) two-tone spectroscopy revels the qubit $f_{01}$ mode as a function of gate voltage. (c) We extract $f_{01}$ as a function of gate voltage from the data, and compare this to simulations of the expected $E_J$ for a given $f_{01}$ to map these values to $E_J$.}
    \label{fig:gatetuning}
\end{figure}

\begin{figure*}[t!]
    \centering
    \includegraphics[width=\textwidth]{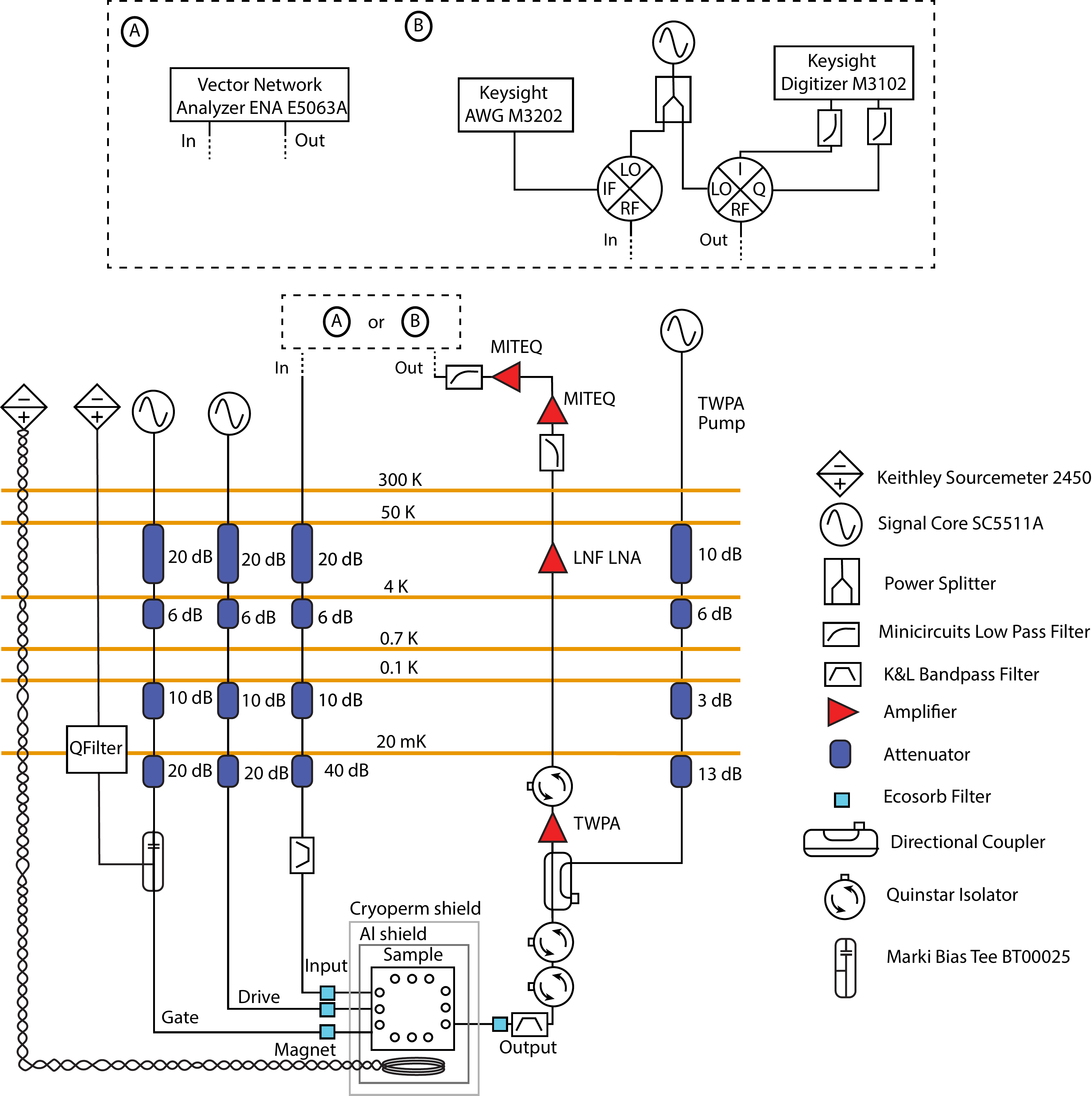}
    \caption{ \textbf{Dilution refrigerator wiring diagram}}
    \label{fig:fridge}
\end{figure*}

\begin{figure*}[t!]
    \centering
    \includegraphics[width=.8\textwidth]{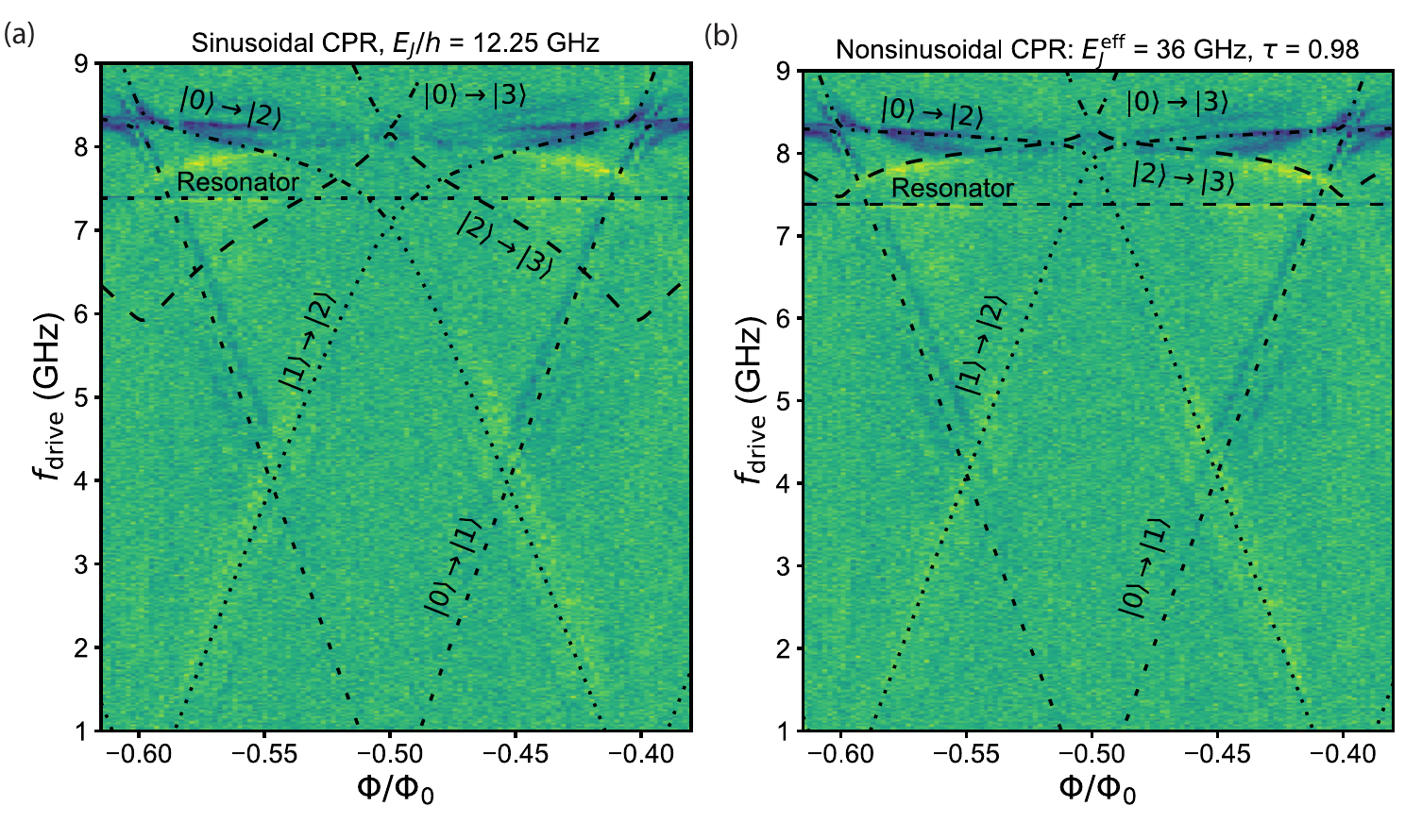}
    \caption{Fits to two-tone spectroscopy data in the heavy regime using (a) a sinusoidal current-phase relation and (b) a nonsinusoidal current-phase relation. In both fits the inductive energy was $E_L/h$ = 2.8 GHz, and the charging energy was $E_C/h$ = .80 GHz. Using the sinusoidal current-phase relation, the Josephson energy is $E_J/h$ = 12 GHz. With the nonsinusoidal current-phase relation, the effective Josephson energy is $E_J^\mathrm{eff}$ = 36 GHz and the transparency is $\tau$ = 0.98. We plot transitions to the first, second, and third excited states, as well as transitions between excited states. The resonator frequency is also shown.}
    \label{fig:fit_comp}
\end{figure*}

\section*{Appendix A: Gatemonium coherence times}
\label{coherencetimes}

The coherence time of a fluxonium qubit depends on the energies $E_J$, $E_C$, and $E_L$. For a fixed set of energies one can estimate the coherence time using Fermi’s golden rule. Using a similar model to that in Ref. \citenum{hazard2019} the decoherence rates of a fluxonium qubit can be written in terms of inductive and capacitive dissipation rates. The inductive loss term is 
\begin{equation}
\Gamma_\mathrm{ind} = \frac{E_L}{\hbar Q_L} \left[ \coth\left( \frac{\hbar \omega_{01}}{2 k_B T}\right)+1\right] |\langle 0|\hat\varphi|1\rangle|^2,
\label{eqn:inductive}
\end{equation}
while the capacitive loss term is 
\begin{equation}
\Gamma_\mathrm{cap} = \frac{\hbar \omega_{01}^2}{8E_C Q_C} \left[ \coth\left( \frac{\hbar \omega_{01}}{2 k_B T}\right)+1\right] |\langle 0|\hat\varphi|1\rangle|^2,
\label{eqn:capacitive}
\end{equation}for qubit frequency $\omega_{01}$, temperature $T$, and inductive and capacitive quality factors $Q_{L}$ and $Q_C$ respectively. In both equations the ground to first excited state transition matrix element for the phase operator $\hat \varphi$ is present. The energy relaxation time is given by $T_1 = (\Gamma_\mathrm{ind} +\Gamma_\mathrm{cap})^{-1}$.

One can see that inductive dissipation is proportional to $E_L$ and capacitive dissipation is inversely proportional to $E_C$. The dissipation rates are also inversely proportional to the capacitive and inductive quality factors. The gatemonium operating frequency is tunable, but we will focus on the heavy gatemonium at half flux, where the operating frequency $\omega_{01}/2\pi$ = 65 MHz. The gatemonium frequency is much lower than that of gatemon qubits and should decrease capacitive dissipation. In addition, the larger charging energy $E_C/h$ = 1 GHz in gatemonium leads to further decreased capacitive dissipation, as opposed to gatemons with charging energies typically between 200 and 500 MHz. We have previously shown lower bounds on the capacitive quality factor to be in the range of $\SI{5.0e4}{}$ by measuring coplanar waveguide resonators, and a similar result was seen in Ref. \citenum{Casparis2018} in a gatemon qubit by measuring $T_1$ over a large frequency range. Assuming gatemonium has a similar $Q_C$ to gatemon qubits, we calculate the expected capacitive dissipation rate assuming $E_L$ = 0.40 GHz, a temperature of $T = \SI{12}{\milli K}$,  and calculating $\langle 0|\hat \varphi|1\rangle = 2.9$ for fluxonium wavefunctions at half flux with this set of energies. We calculate $\Gamma_\mathrm{cap}$ = 7.9 kHz, corresponding to $T_1$ due entirely to capacitive dissipation $T_1$ = \SI{120}{\micro s}, an improvement of more than two orders of magnitude over gatemon $T_1$.

We plot the expected coherence times of devices over the full flux tuning range (between $\Phi/\Phi_0 = 0.0$ to $\Phi/\Phi_0 = 0.5$) for four different values of array inductive quality factor $Q_L$ = \SI{e3}{}, \SI{e4}{}, \SI{e5}{} and \SI{e6}{}, shown in Fig. \ref{fig:t1_estimate}(a). We can see over the whole flux tuning range, $T_1$ is dominated by inductive loss for $Q_L \leq 10^4$. For $Q_L \geq 10^5$, $T_1$ is dominated by inductive loss at low frequency and by capacitive loss at high frequency. It can be seen that coherence times at the half flux sweet spot are quite low, limited by inductive loss in the array to below \SI{10}{\micro s}. However, there is a very large range of flux values which yield substantially increased $T_1$ compared to conventional gatemon qubits. The energy scales here were arbitrarily chosen as to match conventional fluxonium values, and bit-flip times at half flux could be further increased by, for example, decreasing $E_L$. 

To estimate $Q_L$ we again use Fermi's golden rule and our measurements of $T_1$ in Device B. From Fig. 5(b) inset of the main text, we find $T_1$ for Device B to be 78 ns. We would like to calculate the equivalent $Q_L$ given this measurement and an estimate of $Q_C$. In CPW devices with thick Al, which was shown to minimize the inductive loss \cite{strickland2024}, and blanket deposited aluminum oxide, we found a capacitive quality factor $Q_C = \SI{1.2e4}{}$, limited by the blanket gate dielectric. Lifting off the gate dielectric should increase this to $Q_C = \SI{6e4}{}$, which we believe is the limit set by bulk loss in the InP substrate \cite{strickland2024, Casparis2018}. For the energy scales found for Device B, being $E_J/h =\SI{30}{\giga Hz}$, $E_C/h =\SI{150}{\mega Hz}$, and $E_L/h =\SI{4.97}{\giga Hz}$, we calculate the expected frequency, being $\omega_{01}/2\pi = \SI{6.35}{\giga Hz}$, and the calculated $\langle 0|\hat \varphi|1\rangle = 0.3$ matrix element at $\Phi/\Phi_0 = 0.0$. Assuming a temperature $T = \SI{12}{\milli K}$ we solve Eqs. 4 and 5 in the main text given and find that $T_1 = \SI{78}{\nano s}$ and $Q_C = \SI{1.2e4}{}$ corresponds to an inductive quality factor of $Q_L \sim \SI{630}{}$. In this regime we find that the capacitive and inductive dissipation rates are $\Gamma_\mathrm{cap} \sim \SI{3}{\mega Hz}$ and $\Gamma_\mathrm{ind} \sim \SI{20}{\mega Hz}$ respectively. 

% In Fig. \ref{fig:t1_estimate}(b) we show the expected $T_1$ at zero flux as a function of $Q_L$ for a frequency $\omega_{01}/2\pi = \SI{6.35}{\giga Hz}$, estimated $Q_C = \SI{1.2e4}{}$, energy scales $E_J/h = \SI{30.0}{\giga Hz}$, $E_C/h = \SI{0.15}{\giga Hz}$, and $E_L/h = \SI{5.0}{\giga Hz}$. 

% To calculate the equivalent $Q_L$ given this measurement, we estimate $Q_C$ from measurements of CPW resonators with thick Al, as to minimize the inductive loss, and blanket deposited aluminum oxide, we found a capacitive quality factor $Q_C = \SI{1.2e4}{}$, limited by the blanket gate dielectric.

% As found in the analysis at the end of Section \ref{section:timedomain}, $T_1=\SI{78}{\nano s}$ corresponds to $Q_L \sim \SI{630}{}$. We find that in this regime the energy relaxation time is limited by inductive loss,  possibly limited by inductive loss in the thin aluminum film. As seen with CPW resonators in Ref. \citenum{strickland2024} it may be possible that the inductive loss decreases as the film thickness increases.

% While it is unclear exactly what the inductive quality factor is of superconductor-semiconductor JJ arrays, we plan to study the inductive quality factor of superconductor-semiconductor JJ arrays. There have been few experiments on the coherence times of qubits based on InAs junctions in this parameter regime, however results from Ref. \citenum{larsen2020} detailing an enhancement in $T_1$ in a SQUID-based gatemon qubit are quite promising for the inductive quality factor of semiconductor junctions.

\section*{Appendix B: Gate voltage tuning of the Josephson energy}
\label{gatetuning}
The Fermi level in an InAs two-dimensional electron gas (2DEG) can be biased by an applied gate voltage, tuning the number and transparencies of current carrying Andreev bound states (ABS) \cite{beenaker1992}. The inductance and qubit frequency are then sensitive to the applied gate voltage: for decreasing gate voltage, the critical current decreases, the inductance increases, and the qubit frequency decreases. We show single and two-tone spectroscopy as a function of gate voltage for Device A in Fig. \ref{fig:gatetuning}. We see in Fig. \ref{fig:gatetuning}(a) that as the gate voltage is tuned to negative values, the resonator exhibits an avoided level crossing with the qubit. The minimum detuning of these two modes yields twice the coupling strength.

In the same gate voltage range we conduct two-tone spectroscopy, shown in Fig. \ref{fig:gatetuning}(b). We find that for a junction near depletion the qubit undergoes mesoscopic conductance fluctuations, leading to a nonmonotonic behavior with gate voltage \cite{kringhoj2018, danilenko2022}. The plasmon frequency as a function of gate voltage is seen to tune between 9.5 GHz and 5 GHz from -8.5 to -9.4 V. Extracting the qubit frequency from the two-tone data allows us to determine $E_J$ as a function of gate voltage. At zero flux, we are able to calculate the expected $f_{01}$ of a fluxonium with a sinusoidal CPR. The results are then matched to our measured data for $f_{01}$ at zero flux in order to map the measured qubit frequency to an $E_J$ value. We note that while in transmons the plasmon mode should have a frequency of $\sqrt{8E_JE_C}$, due to hybridization of the plasmon and fluxon modes the resulting frequency at zero flux deviates from this.    

\section*{Appendix C: Materials Growth and Fabrication}
\label{fab}
The device is based on an InAs 2DEG grown by molecular beam epitaxy capped with an aluminum layer \textit{in-situ}. Details of the growth procedure can be found in Refs. \citenum{Shabani2016, Kaushini2018, strickland2022}. The structure is grown on a semi-insulating, Fe-doped,  \SI{500}{\micro m} thick, \SI{2}{}-inch diameter, single-side polished InP wafer (AXT Inc.). The oxide is thermally desorbed under an arsenic overpressure in an ultrahigh vacuum chamber. A superlattice and graded buffer layer are grown in order to minimize compressive strain on the active region. The quantum well is formed by a \SI{4}{\nano m} bottom In$_{0.81}$Ga$_{0.19}$As barrier, a \SI{4}{\nano m} InAs layer, and a \SI{10}{\nano m} In$_{0.81}$Ga$_{0.19}$As top barrier. The wafer is delta-doped \SI{6}{\nano m} below the active region with Si. The wafer is then cooled to below $\SI{0}{\degree C}$ for the deposition of a \SI{30}{\nano m} thick Al layer, measured by atomic force microscopy.

The device is fabricated using standard electron beam lithography using polymethyl methacrylate resist. There are three lithography layers. We first dice a $7\times\SI{7}{\milli m ^2}$ piece of the wafer, followed by cleaning in dioxolane and isopropyl alcohol successively. We then define the microwave circuit and perform a chemical wet etch of the Al and III-V layers using Transene Type D and a solution of phosphoric acid, hydrogen peroxide, and deionized water in a volumetric ratio of 1:1:40. We then define the JJ by etching a narrow strip in the aluminum layer, defining two superconducting leads of the JJ separated by \SI{100}{\nano m}. We deposit a \SI{40}{\nano m} thick layer of AlO$_\mathrm{x}$ at \SI{40}{\degree C} by atomic layer deposition, followed by a \SI{100}{\nano m} thick layer of Al serving as the top gate, which is defined by a liftoff procedure. The resulting device is shown in Fig. \ref{fig:fab}(a-c).

\begin{figure}[h!]
    \centering
    \includegraphics[width=.8\linewidth]{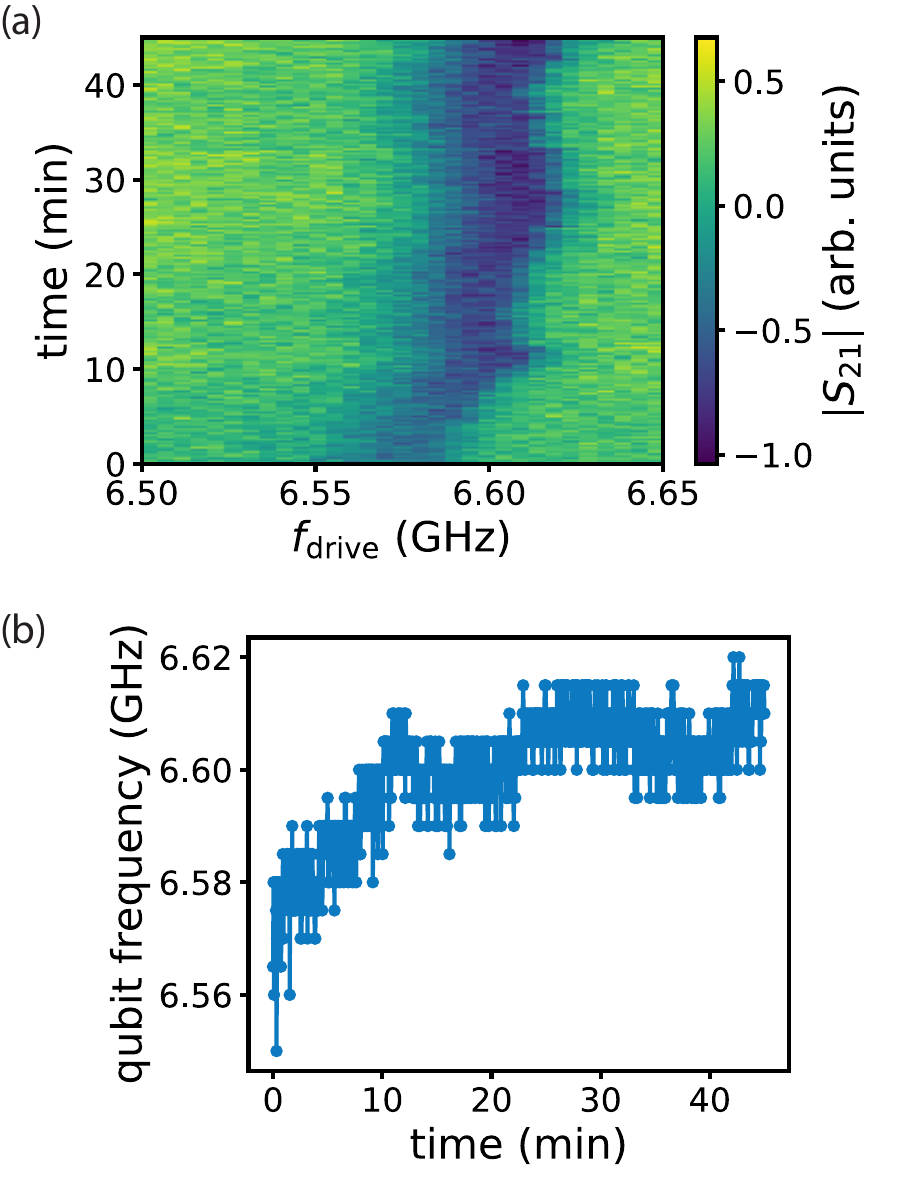}
    \caption{Two-tone spectroscopy of the qubit transition frequency of a similar device over a span of 45 minutes. The magnitude of the complex transmission through a readout resonator $|S_{21}|$ as a function of drive frequency $f_\mathrm{drive}$ is shown in (a) and the extracted qubit frequency over time in (b).}
    \label{fig:timeseries}
\end{figure}

\section*{Appendix D: Fridge Wiring}
\label{wiring}
The chip is mounted on thin copper sample holders and placed in a printed circuit board (PCB). Transmission lines on the chip are connected to waveguides on the PCB by aluminum wirebonds. A gold plated Be/Cu cavity with resonances above 10 GHz encloses the chip, and an Al shield encloses the sample. Magnetic field is provided by a coil within the Al shield. 

The sample is mounted on a cold finger attached to the mixing chamber plate of a cryogen-free dilution refrigerator with a base temperature of 12 mK. Signals are supplied through stainless steel coaxial cables with SMA connectors above the mixing chamber plate, and copper coaxial cables with SMA connectors below the mixing chamber plate. The input rf signal across the common feedline is attenuated by -76 dB from room temperature to base temperature. Drive signals are attenuated by -56 dB. At the mixing chamber plate, the incoming rf signal passes through an Eccosorb filter and a K\&L filter with a dc to 12 GHz pass band. The outgoing signal is passed through another Eccosorb filter and K\&L filter, then to two isolators and a directional coupler before being amplified by a traveling wave parametric amplifier. The tone used to pump the amplifier is attenuated by -39 dB. The signal is then further amplified by a low noise amplifier at 4 K, and then amplified and filtered at room temperature. A voltage source at room temperature supplies dc signals to the chip after being low pass filtered at the mixing chamber plate. dc and rf signals are combined by bias tees at base temperature.

We use a vector network analyzer in order to measure one and two-tone spectroscopy data. Signal generators are used to supply a continuous wave signal to drive the qubit. For time domain measurements, readout and drive continuous wave signals are mixed with the output of an arbitrary waveform generator (AWG) using double balanced mixers. The AWG has a sampling rate of 1 GSa/s. The readout tone is split by a power divider, where one branch is supplied to the local oscillator port of an IQ mixer used to make a readout pulse. The other branch leads to the local oscillator of the readout IQ mixer. The outgoing, amplified signal from the fridge is sent to the rf port, where I and Q quadratures are then recorded using a digitizer with a sampling rate of 500 MSa/s. A schematic of the wiring is shown in Fig. \ref{fig:fridge}.

\begin{figure*}[t!]
    \centering    
    \includegraphics[width=\linewidth]{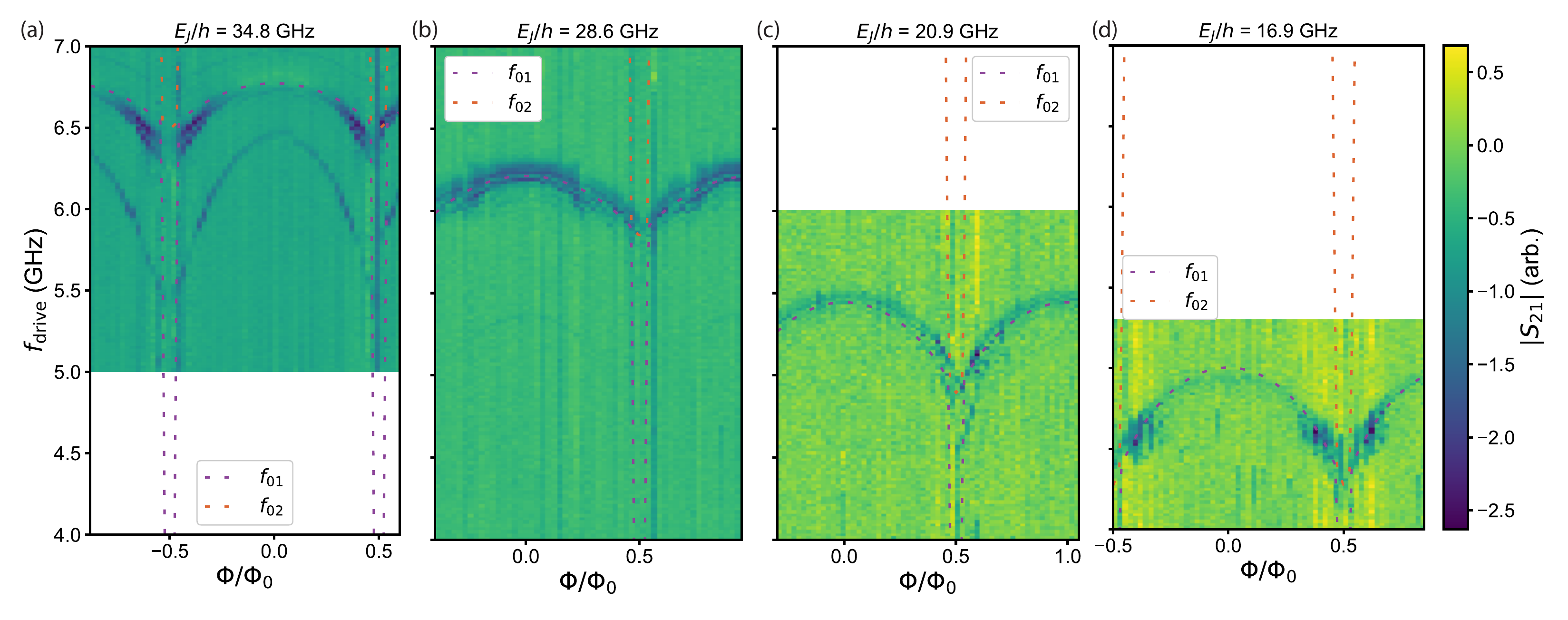}
    \caption{\textbf{Two tone spectroscopy of Device B:} Resonator transmission $|S_{21}|$ as a function of drive frequency $f_\mathrm{drive}$ and applied flux $\Phi/\Phi_0$. The gate voltage tunes the Josephson energy $E_J$ and fits to the data using a sinusoidal current phase relation are shown in dashed lines. It is seen that the dominant features in the data correspond to weakly flux tunable plasmon modes. The theory fits well to the data for $E_J/h$ values of 35 GHz, 29 GHz, 21 GHz, and 17 GHz in panels (a-d) respectively.}
    \label{fig:twotone-B}
\end{figure*}

\section*{Appendix E: Fitting to a nonsinusoidal current phase relation}
\label{nonsinusoidal}

In this section we compare fits to two-tone spectroscopy data using two different models for the qubit transition frequencies using sinusoidal and nonsinusoidal energy-phase relations. As mentioned in the main text, a junction with a nonsinusoidal CPR will have an energy phase relation given by Eq. 2. Assuming the collective modes can be approximated by an single mode with transparency $\tau$ and effective energy scale $E_J^\mathrm{eff}$ we arrive at Eq. 3. A junction with a sinusoidal CPR, typical for Josephson tunnel junctions, have a Josephson energy $E_J(\hat \varphi) = E_J \cos(\hat \varphi)$. The results for fits to two-tone spectroscopy data are shown in Figure \ref{fig:fit_comp}, where the black dotted lines in panel (a) are the transition energies solved from the Hamiltonian with different potentials $E_J(\hat \varphi)$ for each model. For the sinusoidal CPR model in panel (a), we find $E_J$ = 12.25 GHz fits the data best. Results of a nonsinusoidal CPR are shown in panel (b), and we find $E_J^\mathrm{eff}$ = 36 GHz and $\tau$ = 0.98 best fits the data. In both cases it can be seen that the linearly dispersing fluxon transitions corresponding to the $|0\rangle$ to $|1\rangle$ transition are fit well. However, we find that certain features in the data in panel (a) are not fit well, such as the bright yellow transition near $\Phi/\Phi_0$ = -0.57 at 7.5 GHz, and the dark blue transition near $\Phi/\Phi_0$ = -0.43 at 8.0 GHz. Considering the nonsinusoidal CPR, one can see that these features correspond to $|0\rangle$ to $|2\rangle$ and $|2\rangle$ to $|3\rangle$ transitions. Generally, by increasing the transparency make the oscillations of the energy levels with flux flatter near $\Phi/\Phi_0 = 0.0$, and more drastic near $\Phi/\Phi_0$ = 0.5.

\section*{Appendix F: Josephson energy drift over time}
\label{drift}
Charge noise on the active region of a voltage sensitive JJ can lead to unwanted biasing of the current carried by the junction \cite{kringhoj2018, danilenko2022}. We find in our devices that on the order of a few minutes, fluctuations of the qubit frequency occur, presumably from drift in the Josephson energy. We measure the qubit $f_{01}$ transition frequency of a gatemon device every 3 seconds over the course of 45 minutes, as shown in Figure \ref{fig:timeseries}. This gatemon underwent nominally identical fabrication to the gatemonium devices presented here. We find that the qubit frequency undergoes a random walk, characterized by a $1/f$-type noise spectrum. The value of the qubit frequency can change by up to 50 MHz in the span of the measurement, as shown in Figure \ref{fig:timeseries}. At the beginning of the measurement the qubit frequency takes on a value of around 6.56 GHz, and after 45 minutes, the value of the qubit frequency is about 6.61 GHz, a difference of about 50 MHz.

We find that this may lead to instability in the Josephson energy in time in the measurements of our gatemonium device, as seen in Figure 4 (d). We also find that this behavior varies from device to device, and may be sensitive to fabrication and design parameters. Charge noise on the qubit frequency will be directly relevant in gatemon and gatemonium dephasing, and we therefore plan to study this in detail in future work.

\section*{Appendix G: Two-tone spectroscopy of Device B}

In this section we present two-tone spectroscopy data of Device B with fits to a sinusoidal CPR. Parameters of the device are shown in Table \ref{table:1}. The charging energy $E_C/h$ and inductive energy $E_L/h$ are determined to be 150 MHz and 4.97 GHz respectively. The device is in the heavy regime for all measurements shown here. As before, we conduct two-tone spectroscopy by applying a probe tone at a frequency close to that of the readout resonator. Simultaneously, we sweep the frequency of a drive tone, applied through an external charge line which is capacitively coupled to the qubit, driving transitions in the qubit. We plot the transmission of the probe tone across the feedline for varying applied flux and drive frequency. The results can be seen in Fig. \ref{fig:twotone-B}, where four different $E_J$ values can be seen in panels (a) through (d). The most noticeable feature in the data is the presence of a mode which is periodic and weakly tunable with flux, exhibiting a skipping pattern. By plotting the theory on top of the data, for $|0\rangle$ to $|1\rangle$ and $|0\rangle$ to $|2\rangle$ transitions, one can see that these modes correspond to a system of very weakly coupled plasmon and fluxon modes. The frequency of the plasmon mode at zero applied flux tunes down from 6.7 GHz to 5.0 GHz as the gate voltage is depleted. The corresponding Josephson energy also decreases, from 34.8 to 16.9 GHz in this range. At large detunings the dispersive shift becomes less than the linewidth of the resonator. Due to the very large capacitance and relatively small inductance the fluxon modes in this device are not seen in the two-tone spectroscopy measurements.

\end{document}